\documentclass[aps,prc, groupedaddress,superscriptaddress,twocolumn,showpacs]{revtex4}
\usepackage{graphicx}
\usepackage{longtable}
\usepackage{array}
\begin {document} 
\author {S. Mythili}
\affiliation{TRIUMF, 4004 Wesbrook Mall, Vancouver, BC, V6T 2A3, Canada}
\affiliation{Department of Physics and Astronomy, University of British Columbia, Vancouver, BC, Canada}
\author {B. Davids}
\affiliation{TRIUMF, 4004 Wesbrook Mall, Vancouver, BC, V6T 2A3, Canada}
\author {T. K. Alexander}
\affiliation{Deep River, ON, Canada}
\author {G. C. Ball}
\affiliation{TRIUMF, 4004 Wesbrook Mall, Vancouver, BC, V6T 2A3, Canada}
\author {M. Chicoine}
\affiliation{D{\'e}partement de Physique, Universit{\'e} de Montr{\'e}al, Montr{\'e}al, QB, Canada}
\author {R. S. Chakrawarthy}
\author {R. Churchman}
\affiliation{TRIUMF, 4004 Wesbrook Mall, Vancouver, BC, V6T 2A3, Canada}
\author {J. S. Forster}
\affiliation{D{\'e}partement de Physique, Universit{\'e} de Montr{\'e}al, Montr{\'e}al, QB, Canada}
\author {S. Gujrathi}
\affiliation{D{\'e}partement de Physique, Universit{\'e} de Montr{\'e}al, Montr{\'e}al, QB, Canada}
\author {G. Hackman}
\affiliation{TRIUMF, 4004 Wesbrook Mall, Vancouver, BC, V6T 2A3, Canada}
\author{D. Howell}
\affiliation{TRIUMF, 4004 Wesbrook Mall, Vancouver, BC, V6T 2A3, Canada} 
\affiliation{Department of Physics, Simon Fraser University, Burnaby, BC, Canada}
\author {R. Kanungo}
\affiliation{TRIUMF, 4004 Wesbrook Mall, Vancouver, BC, V6T 2A3, Canada} 
\affiliation{Department of Astronomy and Physics, Saint Mary's University, Halifax, NS, Canada}
\author {J. R. Leslie}
\affiliation{Department of Physics, Queen's University, Kingston, ON, Canada}
\author {E. Padilla}
\author {C. J. Pearson}
\author {C. Ruiz}
\author {G. Ruprecht}
\affiliation{TRIUMF, 4004 Wesbrook Mall, Vancouver, BC, V6T 2A3, Canada}
\author{M. A. Schumaker}
\affiliation{Department of Physics, University of Guelph, Guelph, ON, Canada}
\author {I. Tanihata}
\author {C. Vockenhuber}
\author {P. Walden}
\author {S. Yen}
\affiliation{TRIUMF, 4004 Wesbrook Mall, Vancouver, BC, V6T 2A3, Canada}
\date{\today}

\title{Lifetimes of states in ${}^{19}\textrm{Ne}$ above the ${}^{15}\textrm{O}$ + $\alpha$ breakup threshold} 
\begin{abstract}
The ${}^{15}\textrm{O}$($\alpha$,$\gamma$)${}^{19}\textrm{Ne}$ reaction plays a role in the ignition of Type I x-ray bursts on accreting neutron stars. The lifetimes of states in $^{19}$Ne above the $^{15}$O + $\alpha$ threshold of 3.53 MeV are important inputs to calculations of the astrophysical reaction rate. These levels in $^{19}$Ne were populated in the ${}^{3}\textrm{He}$(${}^{20}\textrm{Ne}$,$\alpha$)${}^{19}\textrm{Ne}$ reaction at a $^{20}$Ne beam energy of 34 MeV. The lifetimes of six states above the threshold were measured with the Doppler shift attenuation method (DSAM). The present measurements agree with previous determinations of the lifetimes of these states and in some cases are considerably more precise.
\end{abstract}
\maketitle
 \section{Introduction}
In typical main sequence stars like our Sun, nuclear binding energy is released largely through hydrogen burning via the \textit{pp} chains. In heavier main sequence stars containing carbon, nitrogen and oxygen, the core temperatures are higher and the fusion of hydrogen into helium proceeds mainly through the CNO cycles. These cycles generally dominate over the \textit{pp} chains at temperatures above 20 MK, fusing H into He via a chain of proton captures and $\beta^{+}$ decays while the combined numbers of carbon, nitrogen and oxygen nuclei remain unchanged. When the temperature is lower than $\sim$ 0.1 GK, the cycles are described as cold and the rate of energy release is limited by the slow $^{14}$N($p,\gamma$)$^{15}$O reaction.
   
At higher temperatures, proton capture by $^{14}$N becomes more rapid, resulting in the hot CNO (HCNO) cycles. The rate of energy release in the HCNO cycles is limited by the $\beta^{+}$ decays of $^{14}$O and $^{15}$O, the longest lived radioisotopes in the cycles. At  temperatures  below 0.5 GK, any breakout from the HCNO cycles  is initiated by ${}^{15}\textrm{O}$($\alpha$,$\gamma$)${}^{19}$Ne \cite{Wa81,La86,hahn96,Wi99}. If the rate of $\alpha$ capture by ${}^{15}\textrm{O}$ is comparable to the $\beta^{+}$ decay rate of ${}^{15}\textrm{O}$ in hot, explosive environments, ${}^{19}\textrm{Ne}$ can be formed in appreciable numbers and substantial energy released. The ${}^{19}\textrm{Ne}$ nucleus can then capture a proton, yielding ${}^{20}\textrm{Na}$, which in turn $\beta^{+}$ decays  to ${}^{20}\textrm{Ne}$, breaking out of the HCNO cycles and into the \textit{rp} process. This breakout can lead to the synthesis of nuclei up to ${}^{68}\textrm{Se}$ \cite{Wa97} and beyond \cite{Sc01}. It has been shown  that the rate of the ${}^{15}\textrm{O}$($\alpha$,$\gamma$)${}^{19}\textrm{Ne}$ reaction dramatically affects the behaviour of Type I x-ray bursts on accreting neutron stars \cite{Fi06}. 

It is believed that non-resonant $\alpha$ capture by ${}^{15}\textrm{O}$ does not contribute significantly to the reaction rate at temperatures relevant to accreting neutron stars \cite{La86,Ch92}. Rather, resonant $\alpha$ capture by ${}^{15}\textrm{O}$ leading to states in ${}^{19}\textrm{Ne}$ above the $\alpha$ + ${}^{15}\textrm{O}$ threshold of 3.53 MeV dominates the reaction rate. At temperatures below 2 GK, the ${}^{15}\textrm{O}$($\alpha$,$\gamma$)${}^{19}\textrm{Ne}$ reaction rate depends strongly on a resonance at ${E}_{cm}$ = 500 keV and to a lesser extent on other resonances at  ${E}_{cm}$ =  850 keV and 1070 keV \cite{Ma90}. These resonances correspond to states in ${}^{19}\textrm{Ne}$ at excitation energies of 4035 keV, 4378 keV, and 4602 keV respectively \cite{Ti95,Ta05}.

The decay properties of these important resonances largely determine the reaction rate. The proton and neutron emission thresholds in ${}^{19}\textrm{Ne}$ are 6.4 and 11.6 MeV respectively. The states of interest in the 4-5 MeV region therefore either $\alpha$ or $\gamma$ decay. For a given state the total width $\Gamma$ is then the sum of the $\gamma$ width $\Gamma_{\gamma}$ and the $\alpha$ width $\Gamma_{\alpha}$. The $\alpha$ decay branching ratio and the total width are both required to calculate the resonance strength and the individual contribution of each state to the ${}^{15}\textrm{O}$($\alpha$,$\gamma$)${}^{19}\textrm{Ne}$ reaction rate. Several experiments have been performed to measure the $\alpha$-decay branching ratios of these states \cite{Ma90,La02,Da03,Re03,Vi04,Ta07}. The  total width of each excited state is related to its  lifetime $\tau$ via $\Gamma=\hbar/\tau$. Two measurements of the lifetimes of ${}^{19}$Ne states above the $\alpha$ threshold have been published \cite{Ta05,Ka06}. This paper reports new measurements of the lifetimes of six states in ${}^{19}\textrm{Ne}$ above the $\alpha$ threshold.
 
\section{Experimental setup}
Levels in ${}^{19}\textrm{Ne}$ were populated in the ${}^{3}\textrm{He}$(${}^{20}\textrm{Ne}$,$\alpha$)${}^{19}\textrm{Ne}$ reaction using a 34 MeV $^{20}$Ne beam at the TRIUMF-ISAC facility. To produce the target, 30 keV $^{3}$He ions were implanted at a depth of $\sim$ 0.1 $\mu$m in a 1 cm $\times$ 1 cm $\times$ 12.5 $\mu$m Au foil at the Universit{\'e} de Montr{\'e}al, resulting inÊ a ${}^{3}\textrm{He}$ number density of $\sim 6\times10^{17}$ cm$^{-2}$. Hydrogen, helium and lithium reaction products traversed the entire thickness of the target foil and were detected in a Si surface barrier detector telescope that was centered at 0$^{\circ}$, while the excited ${}^{19}\textrm{Ne}$ ions and the $^{20}$Ne beam were fully stopped in the foil. The maximum velocity of the recoiling ${}^{19}\textrm{Ne}$ ions, which varied slightly with the Q values of the individual resonances, was 0.04\textit{c}, where \textit{c} is the speed of light.

\begin{figure}
\includegraphics[width=\linewidth]{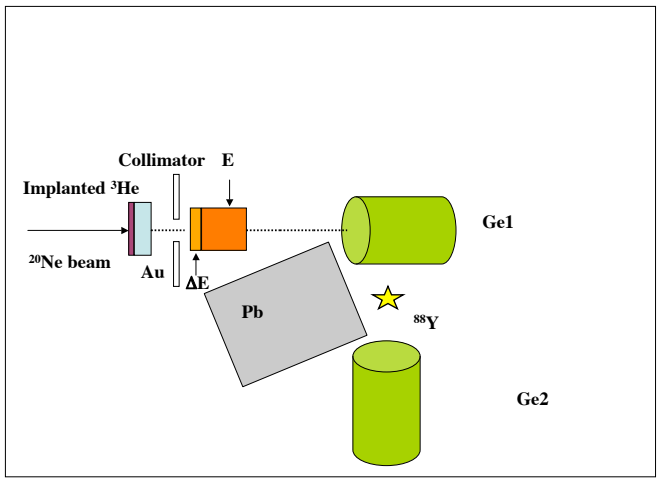}
 \caption{(Color online) The experimental set-up (not to scale). A layer of
${}^{3}$He was implanted in a Au foil. Alpha particles from the
$^3$He($^{20}$Ne,$\alpha$)$^{19}$Ne reaction were detected in 25 $\mu$m
$\Delta$E and 500 $\mu$m E detectors. The Ge detector aligned to the beam
axis was used to detect $^{19}$Ne $\gamma$ rays. The second Ge detector
viewed a weak ${}^{88}$Y source shielded from the target by a Pb absorber.}
 \label{fig1}
 \end{figure}

De-excitation $\gamma$ rays were detected in an 80$\%$ efficient high
purity Ge detector aligned with the beam axis. The line shapes of the
$\gamma$ rays observed with the Ge detector were used to extract the
lifetimes of several states in ${}^{19}\textrm{Ne}$ using the DSAMÊ
\cite{AlRe,Fo74,Fo79,Ke81}. A second Ge detector, which viewed a weak
${}^{88}\textrm{Y}$ source but was shielded from the target, served to
monitor the stability of the electronics. Fig.\ \ref{fig1} is a schematic
representation of the experimental set-up.

A beam of ${}^{20}\textrm{Ne}^{5+}$ was acceleratedÊ to 34 MeV in the
TRIUMF-ISAC linear accelerator system. The average measured current was 75
nA, resulting in an average beam power of approximately 0.5 W. The beam was
focussed to a spot less than 2 mm in diameter, and the beam current was
read off the copper target ladder.

 ${}^{3}\textrm{He}$ is known to remain implanted in Au
foils at room temperature but diffuses out at high temperature. Since
stopping the beam in the foil results in substantial heating, the foil had
to be kept cold to ensure that the ${}^{3}\textrm{He}$ remained in the target.

The scattering chamber holding the target foil contained a copper shroud
coaxial with the beam that was in thermal contact with a liquid nitrogen
reservoir. The shroud was designed to cool the target and surround it
with even colder surfaces during the experiment. This scattering chamber
was used in an earlier experiment \cite{Ka06} and was modified by attaching
additional Cu plates to the Cu shroud, increasing the area of thermal contact with the target ladder and
improving the cooling. The target ladder, when retracted fully, did not make
contact with the cold copper shroud. However, when the target was raised to
the beam axis, thermal contact between the Cu shroud and the target ladder
was made via BeCu fingers attached to a boron nitride plate, cooling the
target below -70$^{\circ}$C. Electrical isolation of the target ladder and
Cu shroud was maintained by the boron nitride plate. The temperature
gradient across the shroud was $\sim$ 330 K/m. This temperature gradient
ensured that the front of the shroud was much colder than the end closer to
the target, favoring the condensation of any hydrocarbons on the colder
shroud and minimizing contamination of the warmer target surface. With a
beam power of 0.5 W, the highest observed temperature of the target ladder
was -58$^{\circ}$C. At these temperatures, very little $^{3}$He diffused
out of the target. The event rate, dominated by $^{20}$Ne + $^3$He elastic
scattering, was monitored constantly during the experiment and served as a
measure of the ${}^{3}\textrm{He}$ concentration in the target.

The scattering chamber had a provision for holding a Si surface barrier
detector telescope. A 25 $\mu$m thick detector served as an energy loss
($\Delta$E) detector and a 500 $\mu$m thick residual
energy (E) detector was placed directly behind the $\Delta$E detector. The
signals from the $\Delta$E detector and the E detector enabled the
identification of particles. The cylindrical detectors each had a surface area of 150
mm$^{2}$ and they were centered on the beam axis. An iron plate with an
aperture was placed between the target and the surface barrier detectors,
defining an angular acceptance of half angle 19$^{\circ}$, corresponding to
a solid angle of 350 msr. The plate allowed two permanent magnets to be placed on the
sides, ensuring the suppression of $\delta$ electrons.
The $\gamma$ rays from states in $^{19}$Ne were detected by the Ge detector
centered on the beam axis 9.5 cm away from the target. The angular
acceptance of the $\gamma$ ray detector wasÊ
$\pm 23^{\circ}$. $\gamma$ rays arriving in coincidence with $\alpha$ particles
were selected for lifetime measurements. 

Following the prescription in Refs.\ \cite{Fo79,Al81}, a second Ge
detector, shielded from the target by 10 cm thick Pb bricks but in view of
a 9 kBq ${}^{88}\textrm{Y}$ source, was used in coincidence with the Ge
detector aligned with the beam axis. Monitoring the 898 keV and 1836 keV
lines from ${}^{88}\textrm{Y}$ allowed a determination of the stability of
the $\gamma$ ray energy measurements during the experiment. The maximum
observed drift was 0.8 keV. Throughout the duration of the run there was a
diurnal variation in the measured $\gamma$ ray energies and the
distribution of the centroids of the source lines could be well
characterized by a Gaussian with a standard deviation of 0.14 keV. 

A coincident detection in both the Si detectors or in both the Ge counters
served as the master trigger. The discriminator signals from the surface
barrier detectors and the Ge counters were also fed into a time-to-digital
converter (TDC), enabling the relative timing of the
signals to be recorded.

\section{Analysis}
\label{analysis}
\begin{figure}
\includegraphics[height=\linewidth, angle = -90]{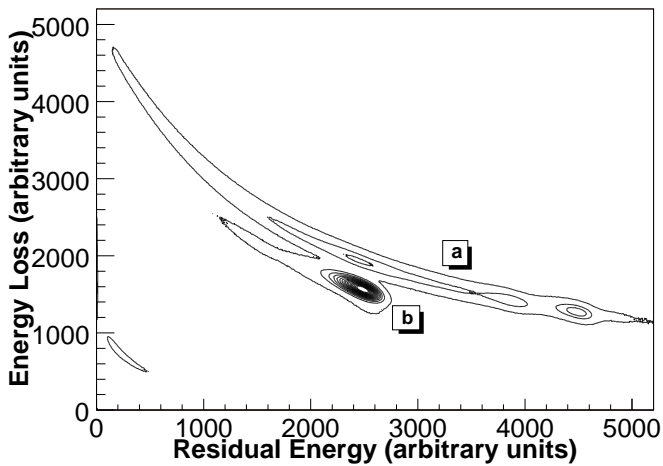}
 \caption{Particle identification spectrum from the Si $\Delta$E and E detectors. The dominant particle groups are (a) $\alpha$ particles and (b) elastically scattered ${}^{3}\textrm{He}$.}
 \label{fig2}
 \end{figure}

\begin{figure}
\includegraphics [width=\linewidth]{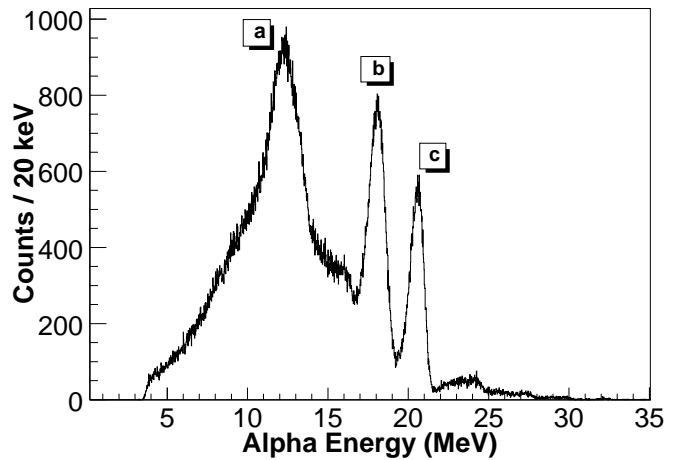}
\caption{Kinetic energy spectrum of detected $\alpha$ particles. The three peaks correspond to the population of $^{19}$Ne states (a) above 4 MeV, (b) around 1.5 MeV, and (c) below 0.3 MeV.}
\label{fig 3}
\end{figure}

Coincident detections in the Si $\Delta$E and E detectors are shown in Fig.\ \ref{fig2}. Using the $\Delta$E and E information shown there, $\alpha$ particles were easily identified and separated from the other particle groups. The total kinetic energy of the $\alpha$ particles was calculated by summing the energy deposited in the $\Delta$E and E detectors. The $\alpha$ particle kinetic energy spectrum is shown in Fig.\ \ref{fig 3}, revealing three prominent peaks corresponding to the population of ${}^{19}\textrm{Ne}$ states with excitation energies above 4 MeV, around 1.5 MeV, and below 300 keV. The levels of interest in ${}^{19}\textrm{Ne}$ above 4 MeV are correlated with $\alpha$ particles having kinetic energies beween 8 and 15.5 MeV. 

\begin{figure}
\includegraphics[width=\linewidth] {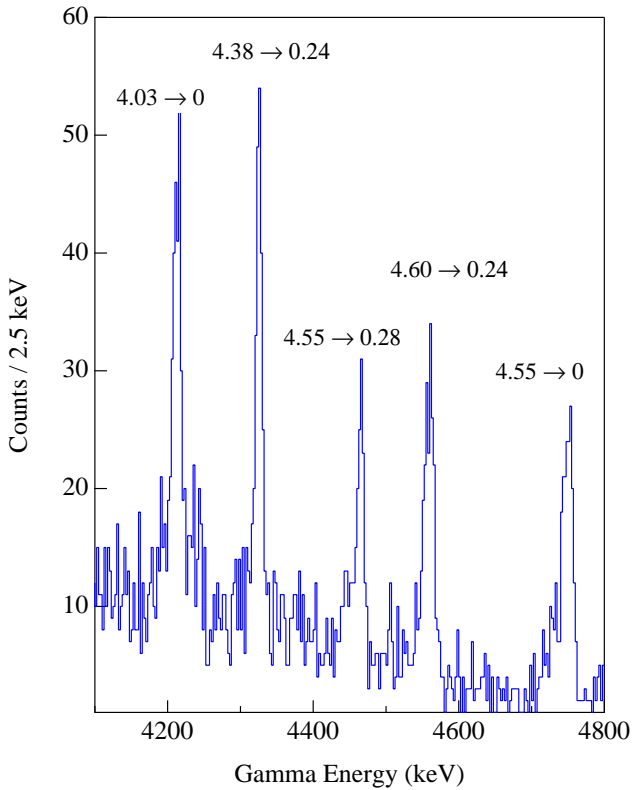}
\caption{(Color online) $\gamma$ rays detected in coincidence with $\alpha$ particles having kinetic energies between 11 and 14.5 MeV. Transitions between states in $^{19}$Ne are labelled by the excitation energies of the states in MeV.}
\label{fig 4}
\end{figure}

The position and width of a timing gate were set by examining the TDC spectrum of Doppler shifted $^{19}$Ne $\gamma$ rays from the decay of the E$_x$ = 4035 keV level to the ground state. Using this timing gate along with a gate on the $\alpha$ particle kinetic energy aided in reducing the background, enhancing the signal to noise ratio. Several transitions with energies above 4 MeV are observed with appropriate $\alpha$ particle kinetic energy and timing gates, as shown in Fig.\ \ref{fig 4}.
 
The $^{19}$Ne level energies from References \cite{Ti95,Ta05} were used to calculate the transition energies. There are seven states in ${}^{19}\textrm{Ne}$  above the ${}^{15}\textrm{O}$ + $\alpha$ breakup threshold whose $\gamma$ decays are well established \cite{Ti95} and all  seven states are known to be directly fed. These states are the 3/2$^{+}$ state at 4035 keV, the (9/2)$^{-}$ state at 4144 keV, the (7/2)$^{-}$  state at 4200 keV, the 7/2$^{+}$ state at 4378 keV, the (1/2, 3/2)$^{-}$ state at 4548 keV, the (5/2$^{+}$) state at 4602 keV and the 13/2$^+$ state at 4634 keV. Doppler shifted lines from all these states were observed in this experiment.

Each of these $\gamma$ rays was analyzed to extract the lifetimes of the decaying states. The $\alpha$ energy gate was narrowed to select $\alpha$ particles with energies between 11 and 14.5 MeV. This enhanced the signal to noise ratio of the $\gamma$ rays originating from the various excited states of ${}^{19}\textrm{Ne}$ further. The line shapes of $\gamma$ rays detected in coincidence with $\alpha$ particles of the appropriate energy and within the relevant timing window were analyzed with the DSAM line shape code described in Ref.\ \cite{Fo79}. This code was developed and used extensively in DSAM experiments at the AECL Chalk River Labs \cite{Fo79,Ke81, Ba82, Al82}.
 
There are three groups of input parameters to the line shape code. These are the geometry of the experimental set-up, the stopping power of the recoiling ions in the target foil, and the detector response parameters.

The code allows the choice of different stopping powers for the $\gamma$ ray emitting recoil in the target foil. We opted for an empirical parametrization of experimental data on the stopping powers of Ne ions in Au \cite{Fo76}. A comparison was made between this parametrization and the stopping powers of SRIM \cite {Zi04}. The stopping powers calculated with SRIM are systematically 5-10$\%$ smaller than the measured stopping powers for all ions heavier than ${}^{7}$Li in Au at the energies of interest. Using SRIM stopping powers would therefore systematically increase the extracted lifetimes.

Another feature of the line shape code is the possibility of using a two foil energy loss calculation. The implantation has the dual effect of decreasing the density of the ${}^{3}\textrm{He}$ + Au layer and increasing its thickness. This results in a change of the stopping power in the implanted region. Following the prescription of Ref. \cite {Al81}, the effect of the implantation on the stopping power was incorporated into the calculation of the lifetimes. In addition to the energy loss of the $^{20}$Ne beam in the front layer of pure Au, the energy losses of the recoiling ${}^{19}\textrm{Ne}$ in the swollen layer of ${}^{3}\textrm{He}$ and Au and in the pure Au layer at the back were taken into account in the line shape analysis.
 
The response of the Ge detector is represented in the third group of input parameters to the code. The detection efficiency and the intrinsic response of the Ge detector are both required. The position, size, and shape of the Ge detector determine its efficiency. The detection efficiency of the detector for 2 MeV and 4 MeV $\gamma$ rays emitted isotropically from the target 9.5 cm away was simulated using a GEANT4 Monte Carlo simulation \cite{Ag03}. The influence of assuming anisotropic dipole and quadrupole angular distributions for the emitted $\gamma$ rays was examined and found to negligibly affect the calculated detection efficiency.

The intrinsic response of the detector was described by a three parameter, skewed Gaussian. Several $\gamma$ rays from a $^{56}\textrm{Co}$ source were analysed with RADWARE \cite{Ra95,Ra95bg} to characterize the intrinsic response of the detector. Since this source has several lines above 2 MeV, the detector response was parameterized for four ${}^{56}\textrm{Co}$ lines, up to a maximum $\gamma$ energy of 3.253 MeV. This parameterization was then linearly extrapolated to the energies of the $\gamma$ rays of interest.

The line shapes predicted for different lifetimes were used in conjunction with a constant background to fit the $\gamma$-ray spectra. The lifetime and the normalization of the background and the Doppler shifted $\gamma$ ray line were varied to obtain the best fit by $\chi^{2}$ minimization. A linear background was also tried, but did not appreciably improve the fits or alter the lifetimes.

Several sources of systematic error in the measured lifetimes were considered. The finite extent of the $^{3}$He layer in the Au foil implies an uncertainty in the location of the reaction vertex that led to systematic uncertainties of up to $\pm$7$\%$ in the lifetimes. The uncertainty in the stopping power and the  beam energy were also found to contribute appreciably to the systematic errors. The uncertainty in the measured stopping power is $\pm$4\% \cite{Fo76}, contributing an uncertainty between $\pm$4\% and $\pm$12\% in the lifetimes. The uncertainty in the beam energy was $\pm$0.2\%, implying an uncertainty that varied with the transition but did not exceed $\pm$11\% of the lifetime values. Other sources of error such as the transition energy uncertainties, variations of the RADWARE fitting parameters, possible misalignments of the Si and Ge detectors, and a shift in the Ge energy calibration were considered and found to contribute negligibly toward the systematic error. Adding the various sources in quadrature, the total systematic uncertainty in the lifetimes measured here ranged from $\pm$4\% to $\pm$19\%.

\section{Results}
This section discusses the line shape analysis of six states in $^{19}$Ne above the $\alpha$ emission threshold that were observed in the experiment. All errors are specified at the 1$\sigma$ level. Where two errors are given, the first is statistical and the second systematic.

\subsection{The 3/2$^+$ state at 1536 keV}
\begin{figure}
\includegraphics[width=\linewidth] {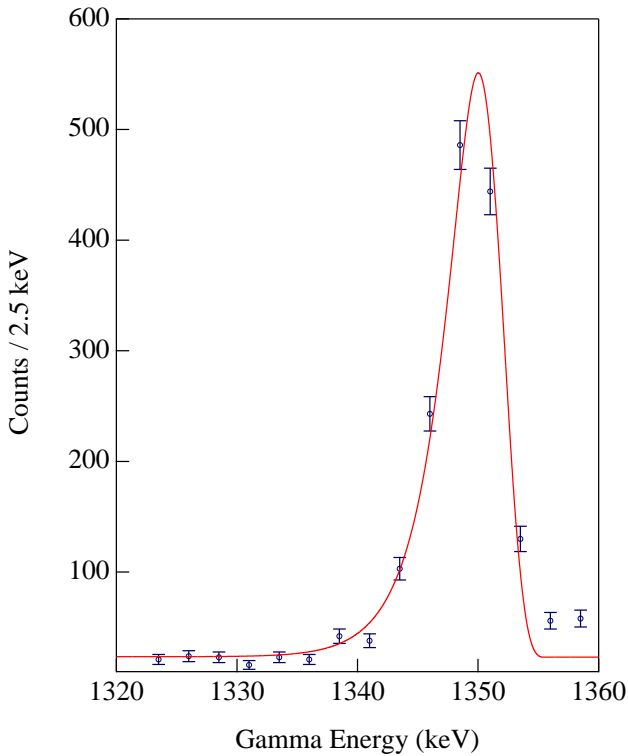}
\caption{(Color online) The Doppler shifted transition of the 1536 keV level to the 238 keV level in $^{19}$Ne. The lifetime of this state was determined to be $19.1^{+0.7}_{-0.6}$ $\pm$ 1.1 fs.}
\label{t1536}
\end{figure}

The input parameters to the line shape code were verified by extracting the lifetime of the well measured state at 1536 keV. The transition of the 1536 keV level to the 238 keV level is shown in Fig.\ \ref{t1536} along with the best fitting line shape. The deduced lifetime of $19.1^{+0.7}_{-0.6}$ $\pm$ 1.1 fs agrees well with the previous measurements \cite{Ti95,Ta05}.

\subsection{The 3/2$^+$ state at 4035 keV} 
This state is predicted to contribute dominantly to the initial breakout from the hot CNO cycles \cite{Fi06}. It has three decay branches \cite{Da73,Ti95}: an $80\pm15\%$ branch to the ground state, a $15\pm5\%$ branch to the level at 1536 keV and a $5\pm5\%$ branch to the level at 275 keV. The two strong transitions from this state were observed in this experiment.

\begin{figure}
\includegraphics[width=\linewidth]{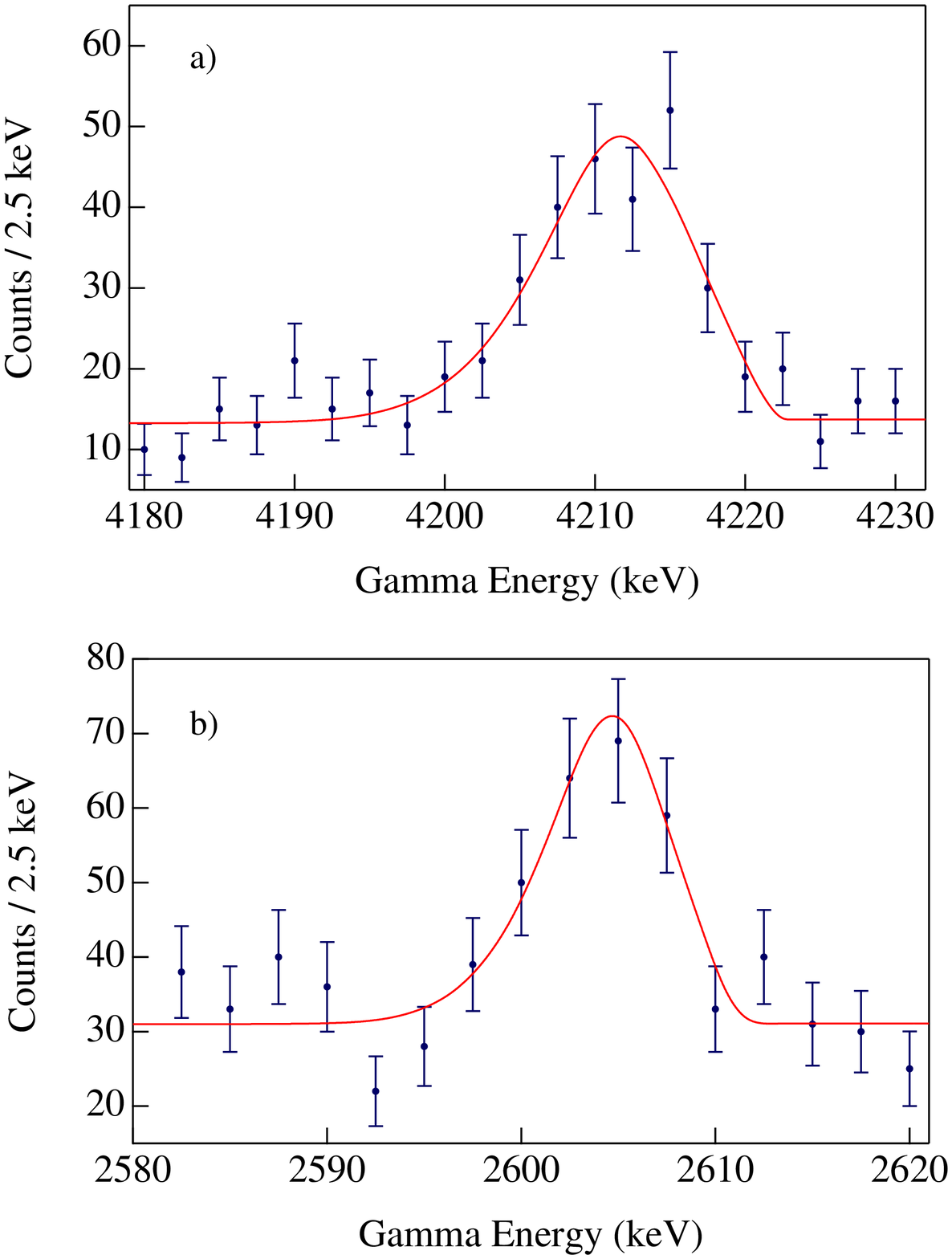} 
\caption{(Color online) Doppler shifted line shapes due to two transitions of the 4035 keV level. The experimental data are shown along with the calculated line shape and background that best fit them. Panel a) shows the decay to the ground state with a lifetime of $7.1^{+1.9}_{-1.9}\pm0.6$ fs. Panel b) depicts the decay to the 1536 keV level with a lifetime of $6.6^{+2.4}_{-2.1}\pm0.7$ fs.} 
\label{fig6} 
\end{figure}

The Doppler shifted ground state transition is shown in Fig.\ \ref{fig 4} and Fig.\ \ref{fig6}(a). Using this transition, the lifetime of this state was found to be $7.1^{+1.9}_{-1.9}\pm0.6$ fs, consistent with the two previous measurements \cite{Ta05, Ka06}. Fig.\ \ref{fig6}(b) shows the second transition of the 4035 keV level, which is discussed next.

\begin{figure}
\includegraphics [width=\linewidth]{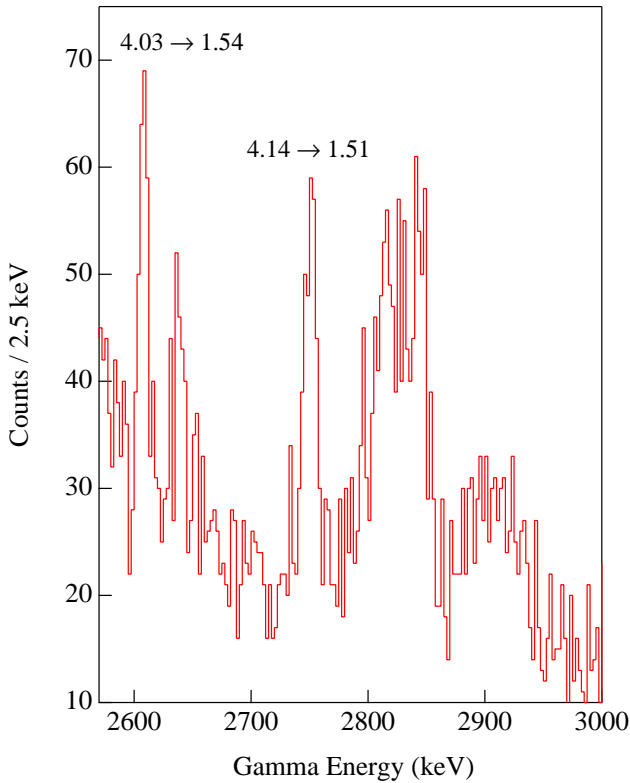}
\caption{(Color online)  $\gamma$ rays detected in coincidence with $\alpha$ particles having kinetic energies between 11 and 14.5 MeV, showing the Doppler shifted $\gamma$ ray due to the transition of the 4035 keV state to the 1536 keV state and that due to the the de-excitation of the 4144 keV level to the 1508 keV state.}
\label{fig5}
\end{figure}

The region between E$_\gamma$ = 2500 keV and 3000 keV shown in Fig.\ \ref{fig5} has several interesting $^{19}$Ne $\gamma$ rays, including the second decay branch of the 4035 keV level. The $\gamma$ ray observed at 2608 keV is the Doppler shifted transition of the 4035 keV level to the 1536 keV level. The transition energy being lower, the background is considerably higher and hence the lifetime determination is not as precise as for the other branch observed at E$_\gamma$ = 4212 keV. The best fitting calculation is shown in Fig.\ \ref{fig6}(b); the lifetime is found to be $6.6^{+2.4}_{-2.1}\pm0.7$ fs, consistent with that obtained from the main decay branch.

\begin{figure}
\includegraphics [width=\linewidth]{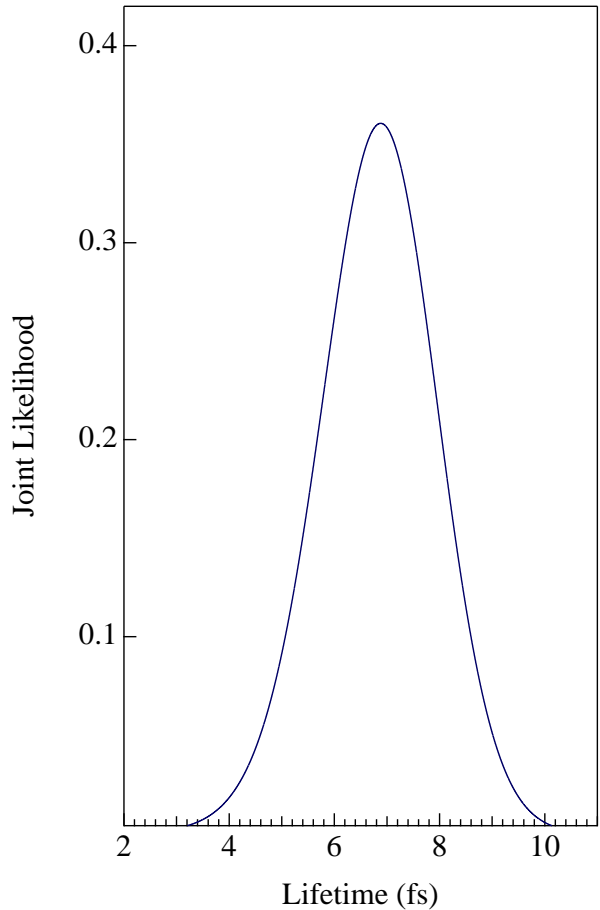}
\caption{(Color online) The joint likelihood for the lifetime of the 4035 keV state in $^{19}$Ne, taking into account the two transitions of the state observed in this experiment. The combined measurements yield $\tau = 6.9^{+1.5}_{-1.5}\pm0.7$ fs.}
\label{fig7}
\end{figure}

Using both transitions, the joint likelihood was computed and the lifetime of the 4035 keV state was found to be $6.9^{+1.5}_{-1.5}\pm0.7$ fs. Fig.\ \ref{fig7} shows a plot of the joint likelihood for the lifetime of this state.

\subsection{The (9/2)$^-$ state at 4144 keV}
\label{section4.2}
\begin{figure}
\includegraphics [width=\linewidth]{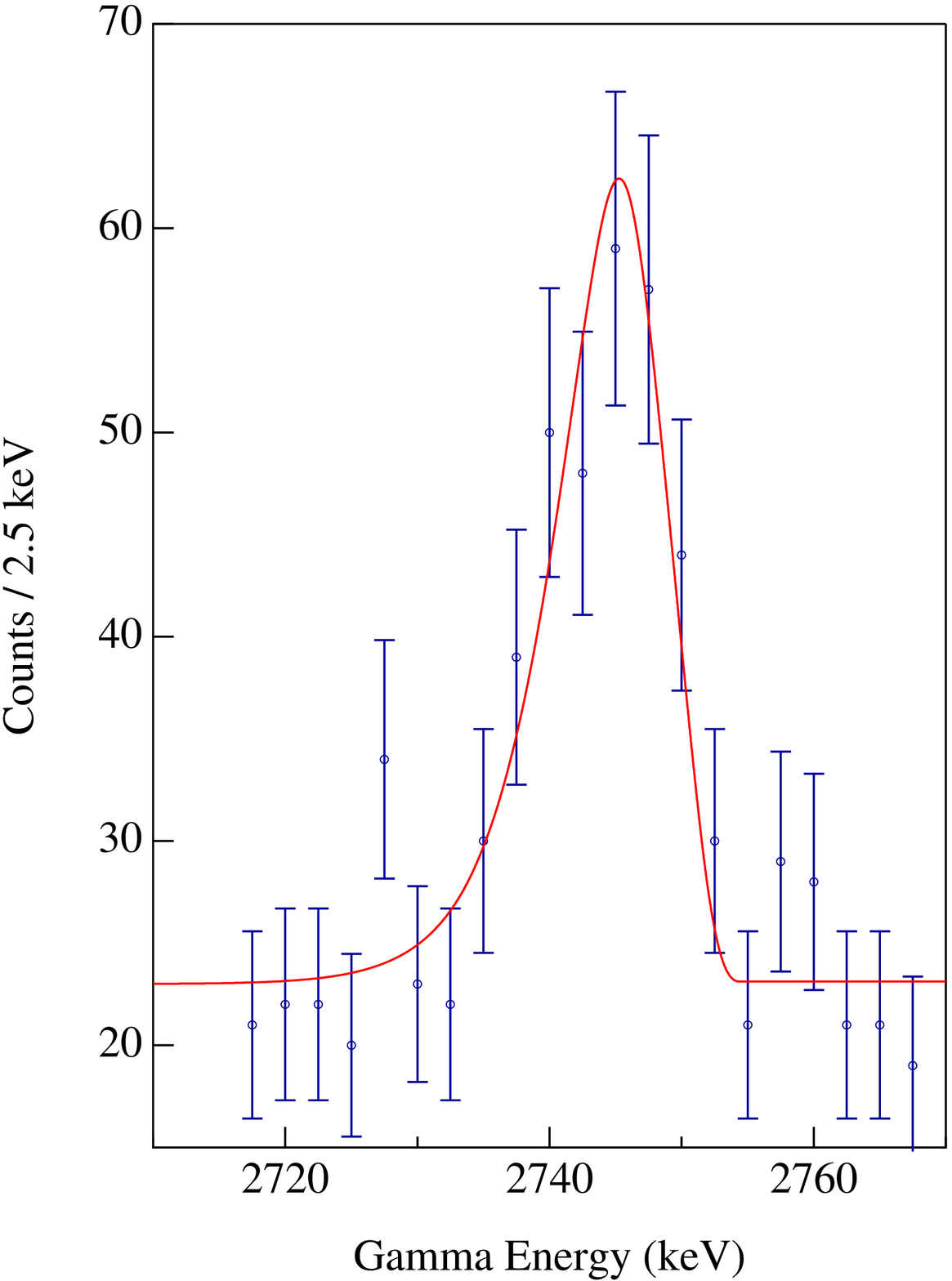}
\caption{(Color online) The $\gamma$ ray transition due to the decay of the state at 4144 keV to the 1508 keV state along with the best fitting line shape calculation corresponding to a lifetime of 14.0 fs.}
\label{fig4144}
\end{figure}

This state decays to the level at 1508 keV \cite{Ti95}. The $\gamma$ ray of 2636 keV is Doppler shifted to yield a line at 2748 keV. This is one of the lines of interest in the E$_\gamma$ = 2500 to 3000 keV region depicted in Fig.\ \ref{fig5}. Applying the appropriate $\alpha$ particle energy and timing gates, this line is  well separated from the other $\gamma$ rays in the region. The lifetime of this level was determined to be $14.0 ^{+4.2}_{-4.0}\pm1.2$ fs. Both the experimental data and the best fitting line shape calculation are shown in Fig.\ \ref{fig4144}.

\subsection{The (7/2)$^-$ state at 4200 keV}

This state is known to decay via two branches: an $80\pm5\%$ branch to the level at 1508 keV  and a $20\pm5\%$ branch to the level at 238 keV \cite{Ti95}. The former transition results in a $\gamma$ ray of 2693 keV while the latter has a higher energy of 3962 keV.

The Doppler shifted 2693 keV transition was identified around 2810 keV, but interference from $\gamma$ rays that were seen with $\alpha$ particles of all energies contributed to a large background underneath and around the line. This large, irreducible background due to fusion evaporation residues limited the precision of the measurement. The lifetime of this level was determined to be $38 ^{+20}_{-10}\pm2$ fs, in agreement with the only other measurement \cite{Ta05}.

\subsection{The 7/2$^+$ state at 4378 keV}

This state is the shortest lived of all the seven states we observed. It decays with $85\pm4\%$ probability to the 238 keV state and with $15\pm4\%$ probability to the state at 2795 keV \cite{Ti95}. This state was the benchmark for testing the Ge detector response. Since the lifetime for this state is short, the predicted line shape was expected to be quite sensitive to the input parameters.

\begin{figure}
\includegraphics [width=\linewidth]{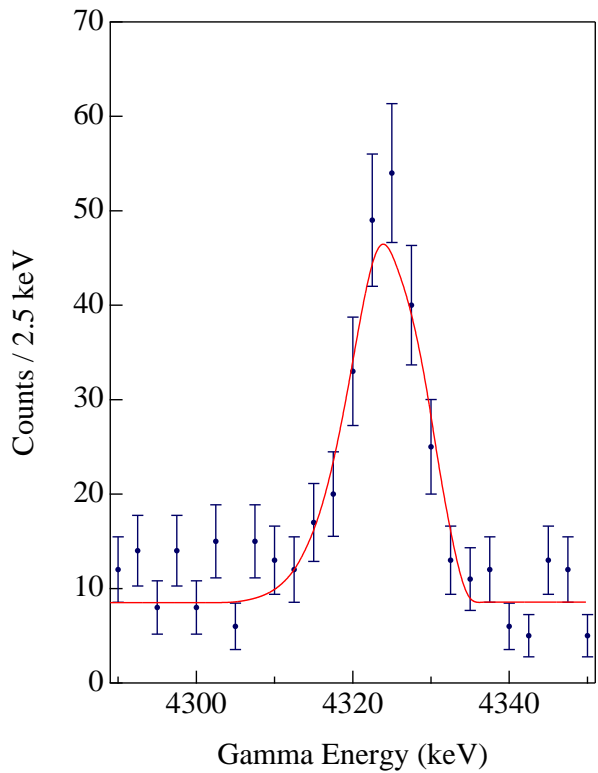}
\caption{(Color online) The $\gamma$ ray transition due to the decay of the state at 4378 keV to the 238 keV state along with the best fitting line shape calculation corresponding to a lifetime of 2.9 fs.}
\label{fig9}
\end{figure}

The stronger of the two transitions was identified with the appropriate $\alpha$ particle energy and timing gates. The line shape was fit and the minimum $\chi^2$ was obtained for a lifetime of 2.9 fs. Fig.\ \ref{fig9} shows the transition and the best fitting calculated line shape. Although it was possible to determine a best fit and upper limits on the lifetime of this state, the statistical power of our data is insufficient to distinguish between the different line shapes predicted for lifetimes shorter than 1.5 fs. For this reason, we do not report a lifetime of the 4378 keV level. In the table we specify instead a 95\% confidence level upper limit on the lifetime of 5.4 fs, which includes systematic uncertainties. The upper 1$\sigma$ statistical and systematic errors were found to be 1.4 fs and 0.6 fs respectively.

\subsection{The (1/2, 3/2)$^-$ state at 4548 keV}

This level decays $65\pm25\%$ of the time to the 275 keV level and $35\pm25\%$ of the time to the ground state \cite{Ti95}. Both transitions were observed and the lifetime of this state was calculated using a joint likelihood approach.

\begin{figure}
\includegraphics[width=\linewidth]{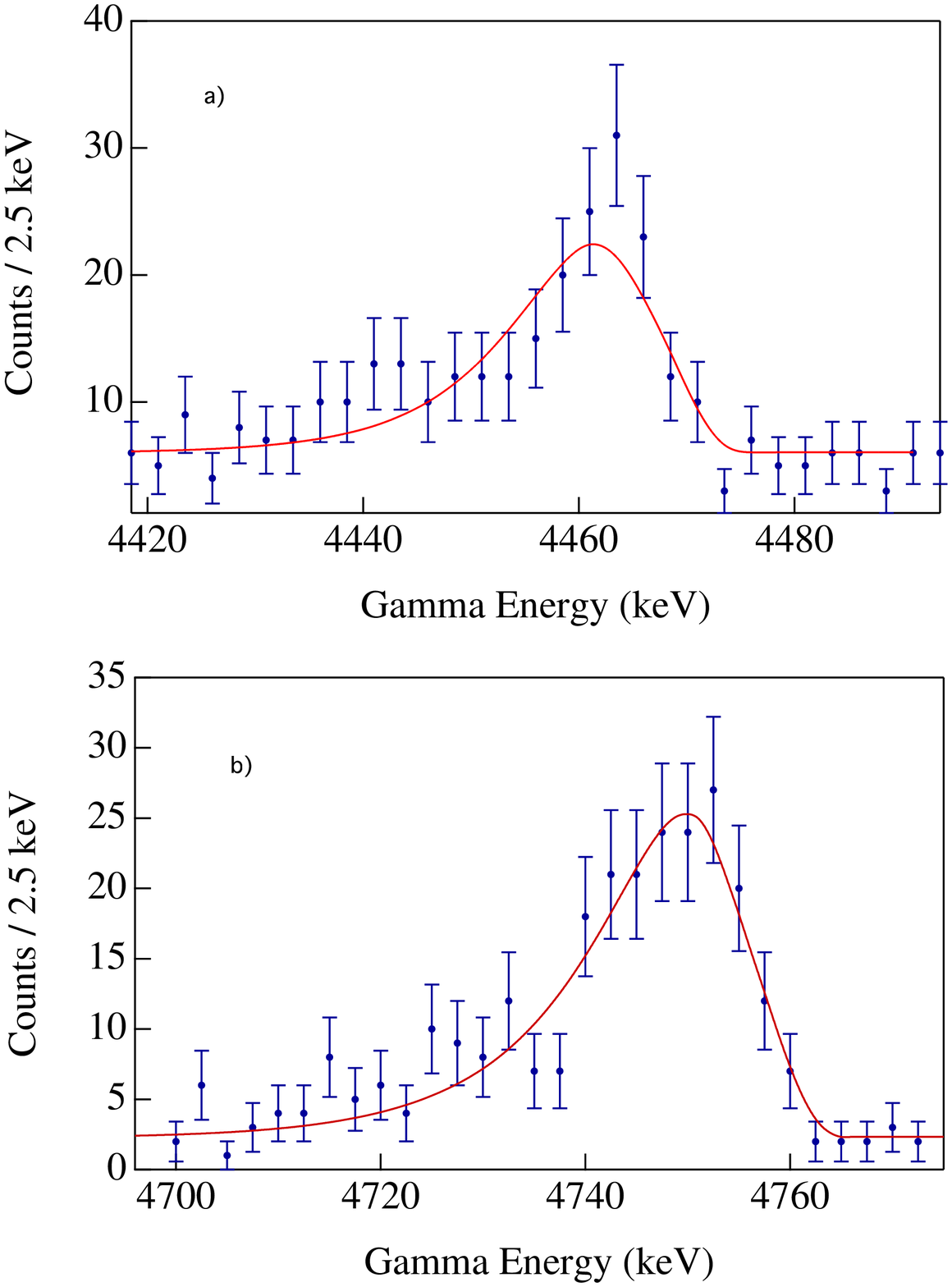} 
\caption{(Color online) Doppler shifted line shapes due to two transitions of the 4548 keV level. The experimental data are shown along with the calculated line shape and background that best fit them. Panel a) shows the decay to the 275 keV state with a lifetime of $16.6^{+4.4}_{-3.6} \pm 1.6$ fs. Panel b) depicts the decay to the ground state with a lifetime of $19.9^{+3.0}_{-3.6}\pm2.3$ fs.} 
\label{l4548} 
\end{figure}

The stronger decay branch has a Doppler shifted line at 4462 keV. The lifetime was determined to be $16.6^{+4.4}_{-3.6}\pm 1.6$ fs using this transition. The data and the best fitting line shape are shown in Fig.\ \ref{l4548}(a).

The $\gamma$ ray corresponding to the transition of the 4548 keV level to the ground state was the highest $\gamma$ ray energy analysed in this experiment. The lifetime deduced from this transition is $19.9^{+3.0}_{-3.6} \pm 2.3$ fs. Fig.\ \ref{l4548}(b) shows the ground state transition and the best fitting calculated line shape.

Both branches yielded consistent values for the lifetime for this state. The most probable value for the lifetime of the 4548 keV state was calculated by combining the lifetimes of the two branches and constructing the joint likelihood, which gives a value of $18.7^{+3.0}_{-2.6}\pm2.2$ fs.

\subsection{The (5/2$^+$) state at 4602 keV}
\begin{figure}
\includegraphics [width=\linewidth]{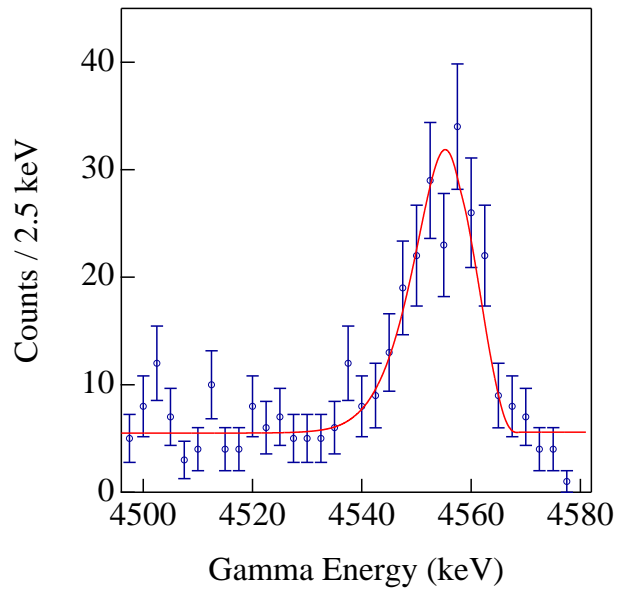}
\caption{(Color online) The $\gamma$ ray transition due to the decay of the state at 4602 keV to the 238 keV state along with the best fitting line shape calculation corresponding to a lifetime of 7.6 fs.}
\label{t4602}
\end{figure}

This state decays via two branches - a $90\pm5\%$ branch to the 238 keV level and a $10\pm5\%$ branch to the 1536 keV level \cite{Ti95}. The Doppler shifted transition to the 238 keV level was observed at 4558 keV with good statistics, allowing the lifetime of this state to be determined. We find $\tau = 7.6^{+2.1}_{-2.0}\pm0.9$ fs. The best fitting calculated line shape is shown along with the data in Fig.\ \ref{t4602}.

\subsection{The 13/2$^+$ state at 4634 keV}
The 13/2$^+$ state at 4634 keV decays exclusively to the 2795 keV state \cite{Ti95}. Its lifetime is expected to be over 1 ps \cite{Ti95} and therefore the observed line shape would reflect a very small shift from the rest frame $\gamma$ ray energy of 1839 keV. Though the 4634 keV state was observed with the appropriate energy and timing gates, random coincidences with the 1836 keV line from our $^{88}$Y source prevented a determination of the  lifetime of the 4634 keV state in $^{19}$Ne.

\subsection{Summary Table}

\begin{figure*}
\includegraphics [width=\linewidth]{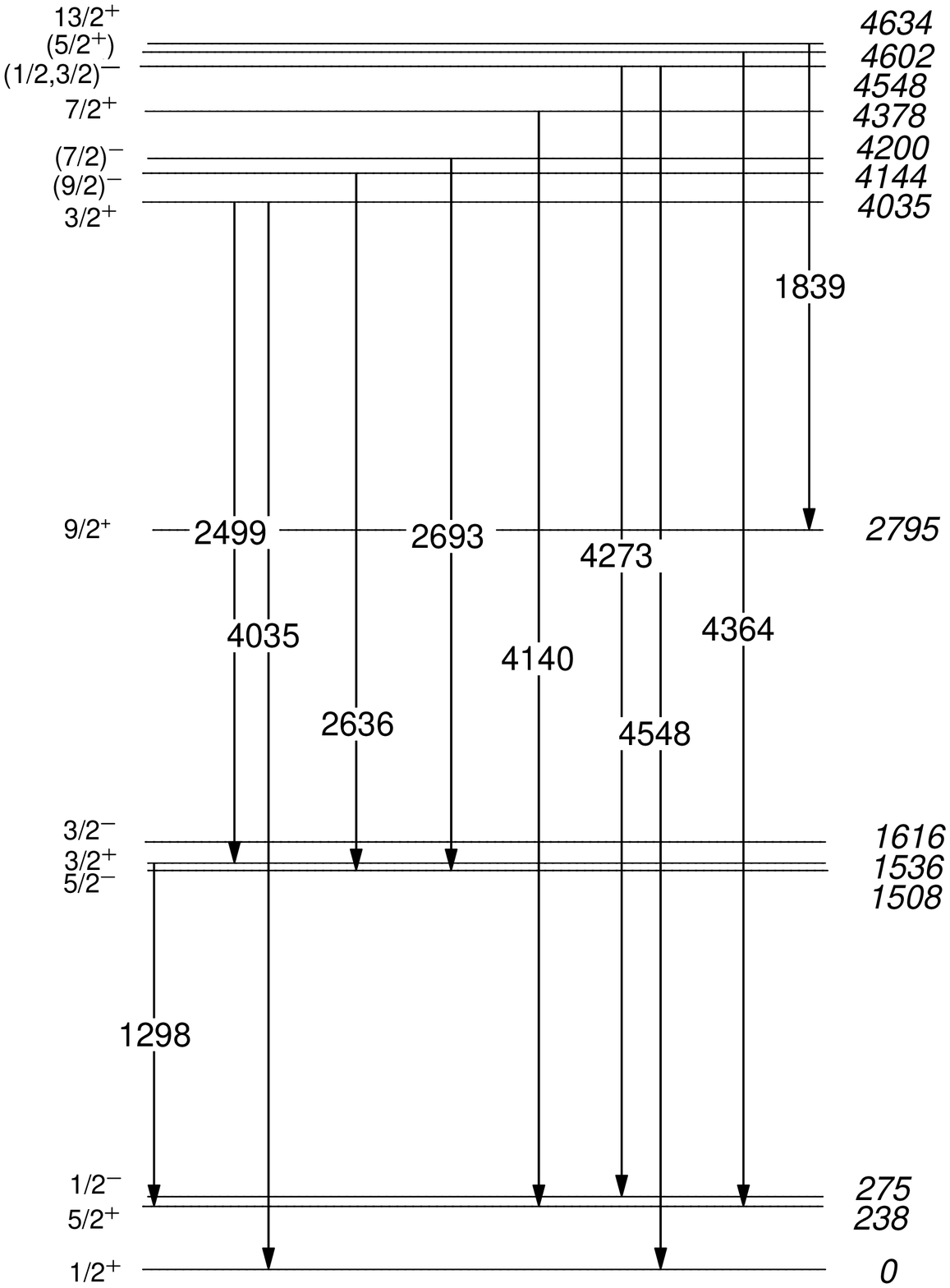}
\caption{Transitions of states in $^{19}$Ne studied in this experiment. The energy levels are labelled by their spins, parities, and excitation energies in keV.}
\label{level}
\end{figure*}

Seven states in $^{19}$Ne above the $^{15}$O + $\alpha$ threshold of 3.53 MeV were populated in this experiment. The lifetimes of six of these states as well as the 1536 keV state in $^{19}$Ne were measured. Fig.\ \ref{level} shows the $^{19}$Ne energy levels and transitions observed in this measurement.

Table \ref{table1} shows a summary of the transitions studied and the lifetimes measured in this experiment. The lifetimes determined here are shown in the rightmost column. The superscript and subscript to the right of each lifetime represent the 1$\sigma$ upper and lower statistical errors and the second number is the 1$\sigma$ systematic error associated with the lifetime. Several sources of systematic uncertainty were considered as described in Section \ref{analysis}. Decays of the 4035 keV and the 4548 keV levels in ${}^{19}$Ne were observed via two branches. The lifetime measurements from both the decay branches were combined to determine the most likely value for $\tau$ and are presented in the table as a combined result.

\section{Discussion}

At astrophysically relevant temperatures, the $^{15}$O($\alpha,\gamma$)$^{19}$Ne reaction rate is dominated by resonant contributions from states in $^{19}$Ne above E$_x$ = 3.53 MeV. The strength of each resonance depends on its $\alpha$ decay branching ratio $B_\alpha$, lifetime $\tau$, and spin $J$ and is given by
\begin{equation}
\omega\gamma=\frac{(2J+1)}{2}B_\alpha(1-B_\alpha)\frac{\hbar}{\tau}.
\end{equation} The lifetimes measured here substantially enhance our knowledge of the astrophysical rate of the $^{15}$O($\alpha,\gamma)^{19}$Ne reaction rate by confirming the previous measurements and improving the precision with which the decay widths of the most important states are known. The implications of the present measurement for the astrophysical reaction rate will be fully discussed in a forthcoming paper. In the future, it would be desirable to better constrain the lifetime of the short-lived 4.38 MeV state. The lifetimes of the other states are now known well enough that they do not introduce significant uncertainties in the reaction rate. 
\begin{table*}
\caption{Lifetimes of states in $^{19}$Ne measured in this experiment and taken from the two previous publications. The 1$\sigma$ statistical and systematic uncertainties determined here are shown separately. The statistical errors are generally asymmetric and are given first as superscripts and  subscripts. The systematic errors are symmetric and follow the statistical errors.}
\begin{ruledtabular}
\begin{tabular}{ccccc}
\hline
Level Energy (keV) &      ${E}_{\gamma}$(keV)  &     Lifetime (fs) & Lifetime (fs)  &  Lifetime (fs)        \\
&Ref.\ \cite{Ti95, Ta05}& Ref.\ \cite{Ta05}& Ref.\ \cite{Ka06} & This work \\
\hline
1536 &    1297.8(4)&        $16\pm4$  &    &   $19.1^{+0.7}_{-0.6}\pm1.1$\\
&&&\\
4035  &    2498.5(9)   &   & &  $6.6^{+2.4}_{-2.1}\pm0.7$\\
&&&\\
     &     4034.5(8)  &   13$^{+9}_{-6}$   & 11$^{+4}_{-3}$ &$7.1^{+1.9}_{-1.9}\pm0.6$\\
&&&\\
     &Combined&&&$6.9^{+1.5}_{-1.5}\pm0.7$ \\
&&&\\
4144 &       2635.9(7)&  $18^{+2}_{-3}$    && $14.0^{+4.2}_{-4.0}\pm1.2$\\
&&&\\
4200  &    2692.7(11)  &   $43^{+12}_{-9}$  & &$38^{+20}_{-10}\pm2$\\
&&&&\\
4378  &      4139.5(6) &   $5^{+3}_{-2}$ & &$\leq5.4$ fs (95\% C.L.)\\
&&&\\
4548   &     4272.6(10)  &   & &$16.6^{+4.4}_{-3.6}\pm1.6$\\
&&&\\
  &     4547.7(10)  &   $15^{+11}_{-5}$  & &$19.9^{+3.0}_{-3.6}\pm2.3$\\
&&&\\
&Combined&&&$18.7^{+3.0}_{-2.6}\pm2.2$\\
&&&\\
4602   &     4363.5(8)&     $7^{+5}_{-4}$ && $7.6^{+2.1}_{-2.0}\pm0.9$\\
&&&\\
\hline
\end{tabular}
\label{table1} 
\end{ruledtabular}
\end{table*}

\section{Summary}

Seven states in  ${}^{19}$Ne above the  $^{15}$O + $\alpha$ threshold were populated in the ${}^{3}$He(${}^{20}$Ne,$\alpha$)${}^{19}$Ne reaction at a beam energy of 34 MeV. The Doppler shifted line shapes of the $\gamma$ decays of these states were analyzed to determine their lifetimes, correcting for density variations due to the ${}^{3}$He implantation in the Au target foil. Two $\gamma$ decay branches of the astrophysically important state at 4035 keV were observed in the experiment and analysis of both branches yielded consistent lifetimes. All of the lifetimes measured here agree with the previous measurements and in some cases considerably improve the precision with which these lifetimes are known. With the possible exception of the 4.38 MeV state, the lifetimes of all the states in $^{19}$Ne relevant to the astrophysical rate of the $^{15}$O($\alpha,\gamma)^{19}$Ne reaction now appear to be known with sufficient precision.

\section{Acknowledgments}
This work was generously supported by the Natural Sciences and Engineering Research Council of Canada. TRIUMF receives federal funding via a contribution agreement through the National Research Council of Canada. SM would like to thank Dave Axen for helpful discussions on GEANT4. 

\bibliography{ne19}

\end{document}